\documentclass[prl,superscriptaddress,twocolumn,floatfix,showpacs,letterpaper]{revtex4}
\usepackage{amsmath}
\usepackage{graphicx}
\usepackage{times}
\usepackage{hyperref}
\usepackage{float}
\usepackage{color}
\floatstyle{plaintop}
\restylefloat{table}
\begin{document}

\title{Prediction of Novel High Pressure H$_2$O-NaCl and Carbon
  Oxide Compounds\\ with Symmetry-Driven Structure Search Algorithm}

\author{Rustin Domingos} 
\affiliation{Department of Physics, University of California, 94720 Berkeley}
\author{Kareemullah M. Shaik}
\affiliation{Department of Earth and Planetary Science, University of California, 94720 Berkeley}
\author{Burkhard Militzer} 
\affiliation{Department of Earth and Planetary Science, University of California, 94720 Berkeley}
\affiliation{Department of Astronomy, University of California, 94720 Berkeley}

\begin{abstract}
  Crystal structure prediction with theoretical methods is
  particularly challenging when unit cells with many atoms need to be
  considered.  Here we employ a symmetry-driven structure search
  (SYDSS) method and combine it with density functional theory (DFT)
  to predict novel crystal structures at high pressure. We sample
  randomly from all 1,506 Wyckoff positions of the 230 space groups to
  generate a set of initial structures. During the subsequent
  structural relaxation with DFT, existing symmetries are preserved,
  but the symmetries and the space group may change as atoms move to
  more symmetric positions. By construction, our algorithm generates
  symmetric structures with high probability without excluding any
  configurations. This improves the search efficiency, especially for
  large cells with 20 atoms or more. We apply our SYDSS algorithm to
  identify stoichiometric (H$_2$O)$_n$-(NaCl)$_m$ and C$_n$O$_m$
  compounds at high pressure.  We predict a novel H$_2$O-NaCl
  structure with \textit{Pnma} symmetry to form at 3.4 Mbar, which is
  within the range of diamond anvil experiments. In addition, we
  predict a novel C$_2$O structure at 19.8 Mbar and C$_4$O structure
  at 44.0 Mbar with \textit{Pbca} and \textit{C2/m} symmetry
  respectively.
\end{abstract}

\maketitle

\section{Introduction}
Crystal structure prediction with theoretical methods is a challenging
subject even for the simplest materials~\cite{Maddox1988,Needs2016}.
For complex materials with a large number of atoms in the unit cells,
$N$, finding the ground state crystal structure is particularly
difficult. The dimensionality of the search space grows as $3N+3$,
while the number of local minima increases exponentially with the
dimensionality, thus the effort required to find the global Gibbs free
energy minimum increases exponentially with $N$. This problem is
classified as NP-hard, and for large systems, searching all
configurations is unfeasible.  However, significant progress has been
made with evolutionary algorithms~\cite{Oganov, Wang2010,Zurek2011},
random search techniques~\cite{AIRSS2006,AIRSS2011}, and others
methods including simulated annealing~\cite{WoodleyCatlow2008}, minima
hopping~\cite{Goedecker2004}, and metadynamics~\cite{Laio2002}. In
principle these methods do not require experimental input, however,
efficiency may be improved if it is incorporated. For example, powder
diffraction data has been used to restrict the search space to within
a known space group~\cite{Meredig2013}. In addition, knowledge
  of energetically preferred structural elements like molecules and
  functional groups may guide the structure generation process. One
  could, for example, place entire H$_2$O molecules or NaCl pairs
  instead of individual atoms. However, one must be aware that these
  geometries may not be preserved at megabar pressures and thus
  the implementation of such
  constraints may eliminate the most favorable structures. While no
search method offers a rigorous path to finding the most stable structure,
they have all lead to many novel low-enthalpy candidate structures
and enriched our understanding of materials at high-pressure. Most
importantly, a number of theoretical predictions have later been
confirmed experimentally
~\cite{Oganov200695,Ono2007,MaLithium2008,Marques2011,Ma2009,Ma2011,Monserrat2016,Pickard2016,Needs2016,Dewaele2016,Struzhkin2016,Reilly2016}.
Knowledge of the globally stable structure allows one to calculate
physical properties under extreme conditions, where experimental
results are not yet obtainable.

For large unit cells, crystal structure prediction has remained a
challenge. When we applied the {\em ab initio} random structure search
technique~\cite{AIRSS2006}, with no symmetry constraints, to look for novel FeSiO${_3}$
structures~\cite{Zhang2014}, we found 74\% of the randomly generated
5-atom cells relaxed into symmetric structures, while the remaining
ones had no (or $P_1$) symmetry. During the relaxation of 10-atom
cells, the fraction of symmetric, non-$P_1$ structures decreased to
43\%. For 15-, 20-, 25- and 40-atom cells, the fraction of non-$P_1$
structures dropped 0.6\%, 0.9\%, 0.01\%, and 0.02\%,
respectively. This means more than 100 20-atom structures needed to be
relaxed in order to generate one symmetric structure that had a chance
of being the global enthalpy minimum. This argument adopts the common
assumption that the most stable structure has at least one symmetry
operation. The wealth of experimental data shows that most compounds
crystallize into symmetric structures at low temperature. This
tendency is expressed by Pauling's {\em rule of
  parsimony}~\cite{Pauling1929} and supported by the energetics of
symmetry calculations~\cite{Wales1998}.

\subsection{Symmetry and Structure Prediction}

It has been recognized that symmetries are key to studying large
clusters~\cite{Wheeler2007,Lv2012,Oakley2013} and crystal
structures~\cite{Wang2010,AIRSS2011,Wang2012,Oganov2013}. In
  Ref.~\cite{AIRSS2011}, this point is addressed by choosing $N_{op}$
  specific symmetry operations. A subset of the atom positions are
  chosen randomly and the remaining images are generated according to
  symmetry. For the example of a mirror plane, the positions for half
  of the atoms are chosen randomly while the other half are placed on
  their mirror images. During the subsequent structural relaxation, a
  more symmetric structure may emerge. With the added symmetry
  operation, this structure can be derived from a supergroup of the
  original group with $N_{op}$ operations. The evolutionary algorithm in
  Refs.~\cite{Wang2010,Wang2012} relies on the particle swarm
  optimization method to move from one generation to then next. The set of
  230 space groups have been used to generate the initial set of
  structures by selecting Wyckoff positions within a given space group
  that are consistent with the chosen composition. Once a structure is
  generated from a particular space group, the algorithm introduces a
  penalty to prevent the generation of another structure from the same
  space group. When this method was applied to the structural
  optimization of TiO$_2$ with classical potentials~\cite{Wang2012} it
  was shown in that the implementation of symmetry constraints
  improved the efficiency of the search algorithm. With constraints,
  more low energy structures were generated and approximately half as
  many generations were needed to find the optimal structure.  In
Ref.~\cite{Oganov2013} symmetry constraints are implemented by placing
atoms on the most general Wyckoff position with the option of merging
nearby atoms onto more symmetric Wyckoff positions, while allowing for
symmetry breaking in subsequent generations. It was shown
in~\cite{Oganov2013} that initializing the evolutionary algorithm with
symmetric structures improved efficiency while using classical
potentials to determine ground state structures of MgAl$_2$O$_4$. 
  In addition, implementation of symmetry constraints allowed the
  determination of the ground state of
  Mg$_{24}$Al$_{16}$Si$_{24}$O$_{96}$, a 160 atom unit cell structure
  that was not found previously without symmetry constraints. In our
approach, we directly sample from the 230 space groups and all
associated 1,506 Wyckoff positions. This allows us to include all
Wyckoff positions consistently and select them with a high probability
without excluding any structure in principle. All space groups and
Wyckoff positions are treated with equal probability until we
eliminate structures in which atoms are very close. The first
water-salt structures that we generated with this method were reported
in Ref.~\cite{Domingos2016}. Independent of this work, a similar
approach was developed in Ref.~\cite{Zurek2017}.  First a set of space
groups is selected. For every space group, a list of all possible
combinations of Wyckoff positions is assembled that are consistent
with the given composition. For large systems, this list may become exceptionally
large. For this reason, the size of this list was reduced by putting
similar Wyckoff positions into groups. This made the algorithm more
efficient but also changed the probability of how often certain
combinations of Wyckoff positions are selected. Conversely, in our algorithm, we
sample Wyckoff positions without generating such a list and
have thus no need to restrict its size. In
Ref.~\cite{Zurek2017}, 10 space groups were chosen when the initial
generation of TiO$_2$ structures were derived for the subsequent
evolutionary algorithm. Using classical potentials, it was shown that
symmetry constraints increase the probability of finding low energy
structures, but also the probability of generating high energy
structures, resulting in an increased average energy overall.

\subsection{High-pressure water-salt and carbon oxides}

Our goal is to design an efficient method to predict the crystal
structure of real materials at arbitrary pressures, without requiring
experimental input. Here we developed a symmetry-driven structure
search (SYDSS) technique to identify novel crystalline compounds at
high pressure. We applied our SYDSS algorithm to search for
(H$_2$O)$_n$-(NaCl)$_m$ and C$_n$O$_m$ compounds at megabar pressures.  While salt
dissolves in water up to a maximum concentration, to our knowledge, no
stoichiometric H$_2$O-NaCl compound has been found in nature,
generated with laboratory experiments, or predicted theoretically.
However, at high pressure, the properties of materials change and
compounds that, while immiscible at ambient conditions, may form
stoichiometric compounds~\cite{FengHennig2008}.  In
Ref.~\cite{Saitta2009}, a novel LiCl$\cdot$6H$_2$O structure was shown to
form at 2 GPa and it was suggested that other salt-ice compounds may
exists at higher pressure.
The discovery of novel high pressure compounds may improve our
understanding of the interior structure and dynamics of ice giant
planets~\cite{WilsonWongMilitzer2013}. If we assume, as an example,
NaCl were available in sufficient quantities, a separate H$_2$O-NaCl
layer~\cite{WilsonMilitzer2014} would form below the ice layer because
of its higher density. The density contrast of the two layers would
also introduce a convective barrier into the interior and potentially
prolong the cooling process of an ice giant planet. 

The properties of carbon and oxygen are of high interest in planetary
science because together with hydrogen and nitrogen, they form the
planetary ices H$_2$O, CH$_4$, and NH$_3$ that make up the bulk of the
interiors of ice giant
planets~\cite{WilsonWongMilitzer2013}. Depending on the formation
conditions and the composition of the building materials, a variety of
planets and different interior structures are expected to
form~\cite{WilsonMilitzer2014, Bond2010,Madhusudhan2012}. Terrestrial
planets like Venus have thick and hot atmospheres that are rich in
CO$_2$. In the atmospheres of more massive exoplanets, we can expect
to find carbon-oxygen compounds that are exposed to yet higher
pressure. However, the properties of such compounds are not yet well
characterized at extreme conditions. With density functional molecular
dynamics simulations, Boates et al.~\cite{Boates2011} predicted CO$_2$
to exhibit a liquid-liquid phase transition at 0.5 Mbar. Leonhardi and
Militzer~\cite{Leonhardi2017} predicted a similar phase transition for
CO to occur between 0.1 and 0.2 Mbar. In the simulations, CO was also
observed to change phase from a molecular to a polymeric fluid. At yet
higher pressures, CO was found to spontaneously freeze into an
amorphous solid. Even though amorphous CO$_2$ structures have been
generated with high-pressure laboratory
experiments~\cite{Santoro2006}, one may expect that the amorphous CO
structures seen in the simulations do not correspond to the
thermodynamic ground state and that there is exists at least one
ordered solid CO structure with a lower free energy. Here we thus use
our SYDSS method to look for novel crystalline carbon-oxygen
structures with a carbon-to-oxygen ratio of 1:1 and variety of other
compositions.

In the original formulation, the {\em ab initio} random structure
search technique did not take advantage of crystal
symmetries and worked in the space group \textit{P}$_1$~\cite{AIRSS2006}.
Symmetric structures emerge, however, when atoms move onto Wyckoff
positions during the relaxation. If one wants to start with, and
maintain a certain set of crystal symmetries during the entire search
process, a special handling of the Wyckoff positions is unavoidable as
the following simple example of a mirror plane illustrates.

If an atom is placed exactly on the mirror plane then there exists
only one instance of it, otherwise there are two. Switching
continuously from one case to the other is difficult within the
context of \textit{ab initio} simulations because when one atom moves
closer to the mirror plane, the distance between the atoms becomes
small, repulsive forces become large and \textit{ab initio}
calculations with pseudopotentials typically do not converge. One
could, of course, remove one of the atoms if the distance between the
pair becomes too small but then the remaining structure would no
longer be symmetric. This means, for structural relaxation algorithms
that preserve the symmetry of the mirror plane, one needs to decide at
the very beginning whether the atom is on or off the mirror plane. In
both cases, the atom can still move in the subsequent relaxation and
occupy a more symmetric position.

While for a single mirror plane, only two cases need to be considered,
for a typical space group there exists a series of Wyckoff positions
that all need to be treated separately. Thus we decided to treat then
230 space groups and associated 1,506 Wyckoff
positions~\cite{IntTables,Bilbao} in a consistent fashion. This
  means even screw axis symmetries are included and atoms, that are
  far away from each other, have a higher chance of being placed on
  symmetry positions. It is our goal to construct an algorithm that
does not exclude any structure but drastically increases the
probability that symmetric structures are generated successfully. To
prevent convergence issues in {\em ab initio} calculations, we
exclude, however, structures where atoms are unphysically close. For
this project, we conservatively chose the following minimum distances
between different Na, Cl, H, C and O species:
r$_{\rm NaNa}$ = 1.4,
r$_{\rm NaCl}$ = 1.2, 
r$_{\rm ClCl}$ = 1.4, 
r$_{\rm NaH}$ = 0.8 
r$_{\rm ClH}$ = 0.8, 
r$_{\rm HH}$ = 0.7, 
r$_{\rm NaO}$ = 1.2, 
r$_{\rm ClO}$ = 1.2, 
r$_{\rm OH}$= 0.8, 
r$_{\rm CO}$= 1.1,
r$_{\rm CC}$= 1.2,
and 
r$_{\rm OO}$ = 1.2 \AA.

\section{Method}

Our SYDSS algorithm repeatedly steps through all 230 space groups
until a user-defined number of structures, $N_S$, have been generated
successfully.  For a chosen space group, it selects lattice parameters
and angles at random, applies any constraints of the space group, and
scales the unit cell so that its volume matches a chosen target
volume. Then it builds a list of all atoms to be placed in the cell
(chemical composition times the number of formula units, $N_{FU}$). As
long as this list is not yet exhausted, our code loops over all
Wyckoff positions of the selected space group. Then it loops over all
atom types that have a sufficient number of atoms remaining to fill
all instances of the selected Wyckoff position. It chooses random
values for all free parameters of this position, generates the
coordinates, and checks whether the atoms satisfy all minimum distance
criteria~\cite{supercellEq}. If they do, our algorithm continues to
place the remaining atoms. If it fails to meet the distance criteria
at any point in this process, it discards the current configuration
and continues with the next space group until all $N_S$ initial
structures have been generated.

\begin{figure}[htb]
  \includegraphics[width=.45\textwidth]{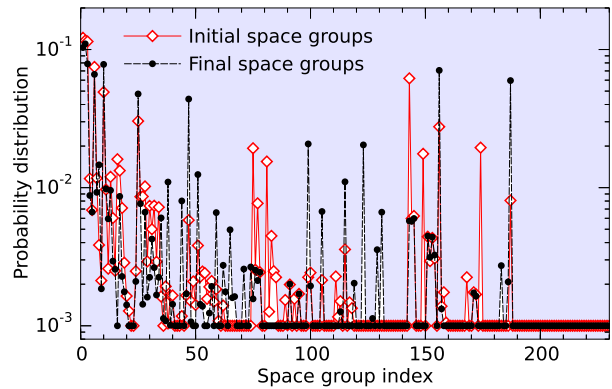}
  \caption{Distribution of initial and final space groups among the
    57,347 H$_2$O-NaCl structures that we generated with our SYDSS
    algorithm. Both distributions are not identical because the space
    group may change during the structural relaxation with DFT forces.
    Some space groups have a low or zero occurrence probability. We set
    their probabilities to 10$^{-3}$ to include them in this graph. }
\label{Fig:initial_final}
\end{figure}

In our current implementation, the SYDSS algorithm has only a minimal
set of adjustable parameters: the atoms in the cell, the set of
minimum distances, and the target volume. The initial unit cell angles
are chosen between 40 and 140 degrees ~\cite{AIRSS2006} because the primitive cells of most structures can be represented in this way. This range could be broadened or one could sample from a smooth prior distribution that includes all angles. These choices, in
particular the distance criteria, imply that not all space groups
occur in the list of generated structures with equal probability.
Figure~\ref{Fig:initial_final} shows the probability distribution of
space groups in the set of initial structures of H$_2$O-NaCl
structures. Many space groups occur with very low or even zero
probability because an insufficient number of atoms remain to fill
all instances of a chosen Wyckoff position, or the inability to do so and satisfy
all minimum distance criteria. In particular, cubic systems (space groups
195-230) occur rarely among our generated H$_2$O-NaCl structures but they occur
frequently when we apply our algorithm to monatomic metals. Because of
its lack of symmetry constraints, space group \textit{P}$_1$ is still
among the space groups that are generated most often but its total
weight is now closer to 10\% compared to 100\% in ~\cite{AIRSS2006}.
If needed, additional biases could be introduced into the current
implementation of our SYDSS algorithm in order to reduce the
\textit{P}$_1$ probability further.

Starting with these initial structures, structural relaxation at
constant pressure was carried out in the framework of density
functional theory, using the Perdew-Burke-Enzerhof
functional~\cite{Perdew1996} and the projector augmented wave method
~\cite{Kresse1996} as implemented in the Vienna \textit{ab initio}
simulation package~\cite{Kresse1999}. A basis-set cutoff energy of 980
eV used for the plane-wave expansion of the wave functions.  For the
first round of relaxation, we used k-point grids of
4$\times$4$\times$4 for cells with less than 15 atoms and
2$\times$2$\times$2 for cells with greater than 15 atoms. The best of
these structures were then re-relaxed using higher density grids
(6$\times$6$\times$6 to 12$\times$12$\times$12) to ensure accurate
enthalpies. This method allowed us to search the structures more
efficiently by removing unlikely candidates early.

\begin{figure}[htb]
  \includegraphics[width=.45\textwidth]{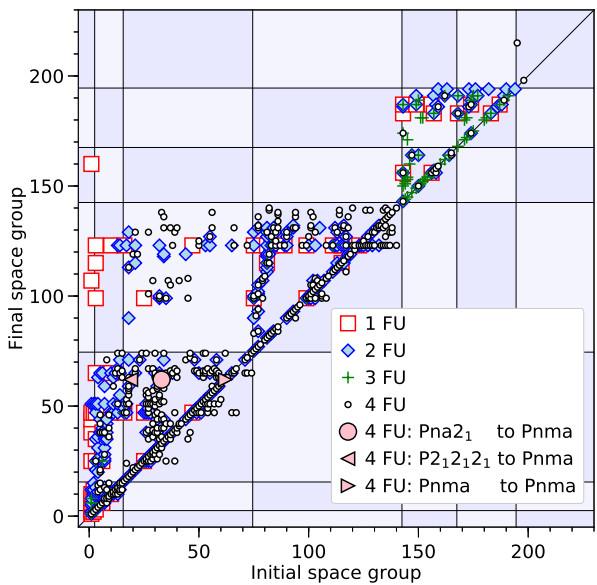}
  \caption{In this transition diagram, final versus initial space groups are plotted for the relaxation of 1 to 4 formula
    unit structures. Transitions to higher space groups are much more
    frequent because the structural relaxation often increases the symmetry. 
    The large circle indicates one possible pathway to our {\it Pnma} structure. The lines separate the 7 crystal systems.}
\label{Fig:transitions}
\end{figure}

During relaxations, the symmetry of the initial space group was
preserved. This still allowed structures to attain higher symmetries
of a supergroup during the relaxation if the atoms move to more
symmetric positions while maintaining the symmetry operations of the
original space group.  This means the space groups of the initial and
the relaxed structures may differ. Such transitions are illustrated in
Fig.~\ref{Fig:transitions}. Many transitions occurred between space
groups in the same crystal system but we also noticed transitions from
monoclinic (space groups 3-15) to orthorhombic (16-74), from
orthorhombic to tetragonal (75-142), and from trigonal (143-167) to
hexagonal (168-194) systems. In a few instances, a smaller primitive
unit cell emerged during the relaxation. Such a transition may plot
below the diagonal in Fig.~\ref{Fig:transitions} because the smaller
unit cell may have a lower space group number.

\section{Results}
\subsection{Water-Salt Structure Search: (H$_2$O)$_n$-(NaCl)$_m$}

We generated and relaxed over 55,000 structures with compositions
(H$_2$O)$_n$-(NaCl)$_n$ having between 1 and 4 formula units at
pressures between 1 and 10 Mbar. We also explored additional
water-salt mixing ratios by relaxing over 11,000
(H$_2$O)$_n$-(NaCl)$_m$ structures with $n$:$m$=4:1, 3:1, 2:1, 3:2,
1:2, and 1:3. However, we were not able to find any thermodynamically
stable structures with $n \ne m$ and we will thus focus the following
discussion on structures with equal water-salt ratios where our
structure search was more successful.

Enthalpies of the computed (H$_2$O)$_n$-(NaCl)$_n$ structures were
then compared with the enthalpy of the H$_2$O and NaCl
endmembers. NaCl endmember enthalpies were calculated using the B2
structure which is stable in this pressure range~\cite{Bassett1968}.
H$_2$O endmember enthalpies were calculated using the high pressure
ice phases predicted at each pressure. At 1-2 Mbar, enthalpies were
calculated using the ice X structure~\cite{Polian1984}. For pressures
of 3-7 Mbar, enthalpies were calculated using the \textit{Pbcm}
structure~\cite{Benoit1996}. At 8 Mbar, the \textit{Pbca}
structure~\cite{Militzer2010} was used. At 9-10 Mbar, the
\textit{P}$3_1$21 structure ~\cite{Pickard2013} was assumed.  To
further test our SYDSS method, we also applied it to pure water
ice. We relaxed 2,000 H$_2$O structures at 9 Mbar and reproduced the
\textit{P}$3_1$21 structure from Ref.~\cite{Pickard2013}.

\begin{figure}[htb]
   \includegraphics[width=.45\textwidth]{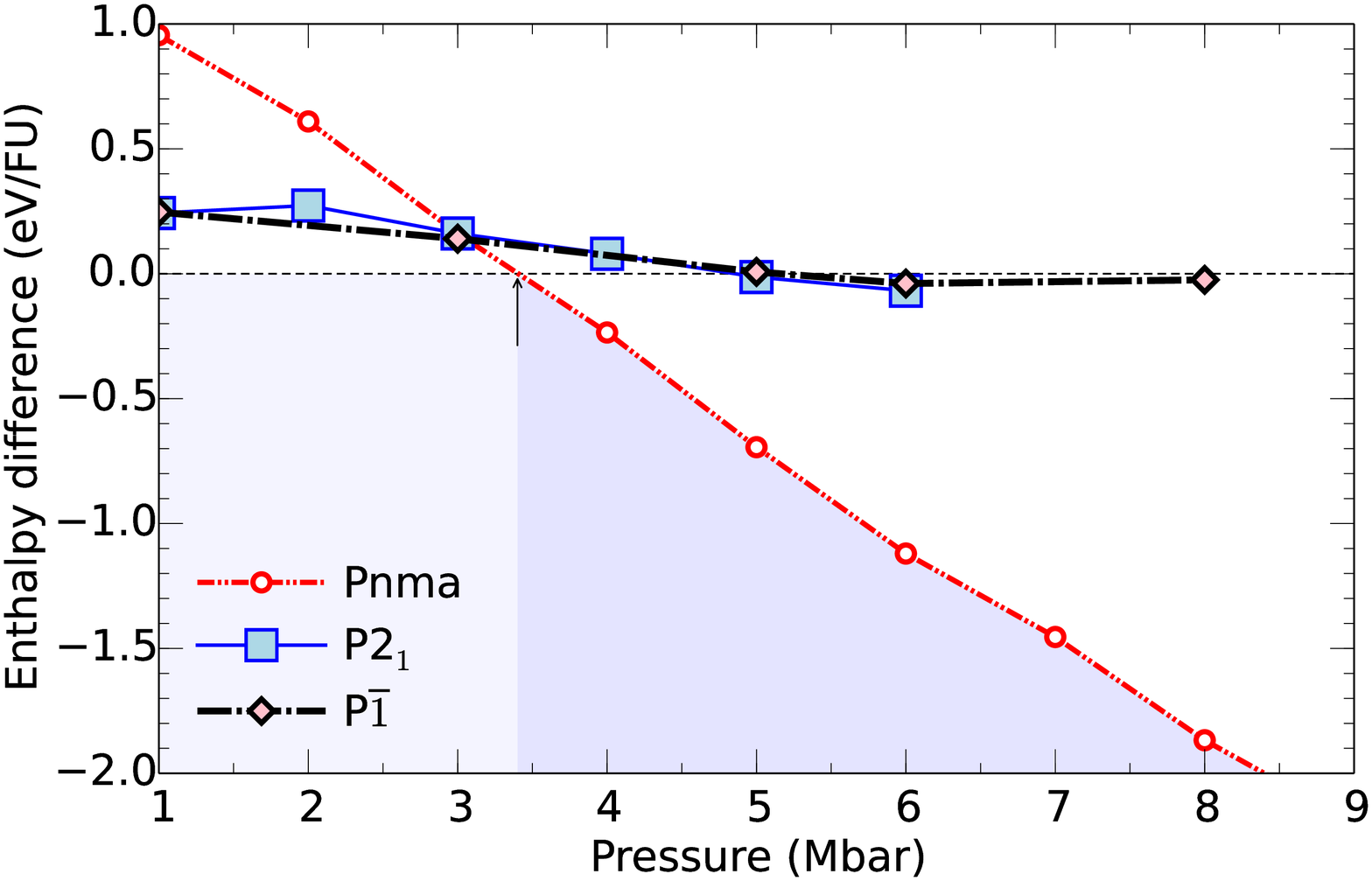}
   \caption{The difference in enthalpy, $H_{\rm H_2ONaCl} - H_{\rm
       H_2O} - H_{\rm NaCl}$, per formula unit as function of
     pressure. The arrow marks the pressure of 3.4 Mbar where the H$_2$O-NaCl
     structure with {\em Pnma} symmetry is predicted to form from
     H$_2$O and NaCl. The $P2_1$ and $P\bar 1$ structures were also
     shown to have lower enthalpies than the endmembers at 5 Mbar but
     the {\em Pnma} structure is energetically favored. }
 \label{Fig:enthalpy}
\end{figure}

After comparison with endmember data, three H$_2$O-NaCl structures
were found to have enthalpies lower than that of the combined
endmembers, suggesting a novel H$_2$O-NaCl structure would form at
high pressure. The enthalpy comparisons of the three best structures
with that of the endmembers is given in Fig.~\ref{Fig:enthalpy}. The
P$\bar{1}$ structure was found by relaxing a structure with one
formula unit of H$_2$O-NaCl, the P$2_1$ structure was found from two
formula unit structures, and the {\it Pnma} structure was found from
four formula unit structures. This enthalpy data predicts a novel
\textit{Pnma} symmetric H$_2$O-NaCl structure forming at 3.4 Mbar,
which is within the pressure range of diamond anvil cell
experiments~\cite{Lobanov2015}.

Out of 7,909 four-formula-unit H$_2$O-NaCl structures that were
successfully relaxed, 185 ($\sim$2\%) relaxed into the \textit{Pnma}
space group. Of these 185 \textit{Pnma} structures, 74 relaxed from
structures with P$2_12_12_1$ symmetry, 35 from \textit{Cc} symmetric
structures, and 76 initially started from \textit{Pnma} symmetric
structures. The fact that this structure was never generated from a
nonsymmetric initial structure (space group $P_1$) and the rate of
occurrence illustrates the advantages of implementing symmetry constraints in our algorithm. 

\begin{figure}[h]
  \includegraphics[width=0.45\textwidth]{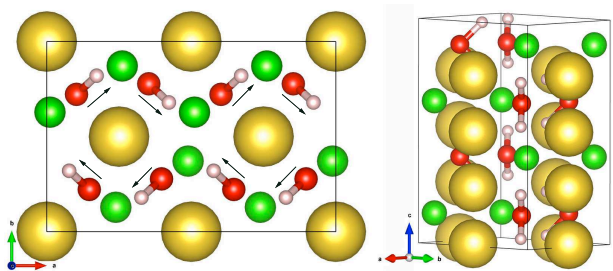}
  \caption{ Novel orthorhombic NaCl-H$_2$O crystal structure with {\it
      Pnma} symmetry. With decreasing size, the spheres denote the positions
    of Cl, Na, O and H atoms. The structure has 20 atoms per unit
    cell but has been doubled in $c$ direction in the right image.}
\label{Fig:images}
\end{figure}

The parameters of our novel orthorhombic H$_2$O-NaCl structure are
given in Tab.~\ref{table1} and two pictures are shown in
Fig.~\ref{Fig:images}. We verified that this
  structure is dynamically stable by performing phonons calculations
  with the Phonopy code~\cite{Togo2015} using 1x1x2, 1x2x1, and 2x1x1
  supercells. The structure can be explained best by
analyzing the layering parallel to the $a$-$b$ planes. Layers with Cl$^-$ ions
alternate with layers of Na$^+$ and O$^{2-}$ ions. The Cl$^-$ ion
always occupy the same position in every layer. The Cl$^-$-Cl$^-$
distances are thus smaller than the separation between other ion pairs
of the same type. From layer to layer, the Na$^+$ and O$^{2-}$ ions
alternate between two positions, leading to unit cell with 20
atoms. The typical geometry of a H$_2$O molecule is well
preserved. The H$_2$O dipole moments lie in the $a$-$b$ planes and are
arranged in clockwise or anticlockwise direction around each column of
Cl$^-$ ion. Overall the charges are reasonably well balanced in this
structure. 

\begin{table}
\begin{tabular}{c c c c c}
\hline
\hline
Atom& Wyckoff & x & y        & z \\
\hline
Na& b & 0       & 0       & 1/2 \\
Cl& c & $-$0.476   & 1/4     & $-$0.370\\
H &  d & ~~0.355   & $-$0.437 & $-$0.310 \\
O &  c & ~~0.278   & 1/4     & ~~0.267 \\
\hline
\hline
\end{tabular}
\caption{Parameters of the orthorhombic H$_2$O-NaCl structure with {\it Pnma}
  symmetry at 4 Mbar. The lattice parameters are $a$=3.942, $b$=3.849, and $c$=5.187 \AA.}
\label{table1}
\end{table}

C

We generated and relaxed over 700,000 C$_n$-O$_m$ structures with up
to 52 atoms per unit cells with ratios from n:m ranging from 1:7 to
6:1 at pressures between 1 and 50 Mbar. Enthalpies of the resulting
structures were compared to the C and O endmembers, which were
calculated using the stable carbon phases of diamond for 1-10 Mbar,
BC8 for 15-25 Mbar, and SC1 for 30-50 Mbar~\cite{Benedict2014}. For
oxygen endmembers $\zeta$-C2/m oxygen for 1-15
Mbar~\cite{Akahama1995,Ma2007} and the \textit{Cmcm} oxygen structure
above 20 Mbar~\cite{Sun2012} were used.

\begin{figure}[h]
  \includegraphics[width=.45\textwidth]{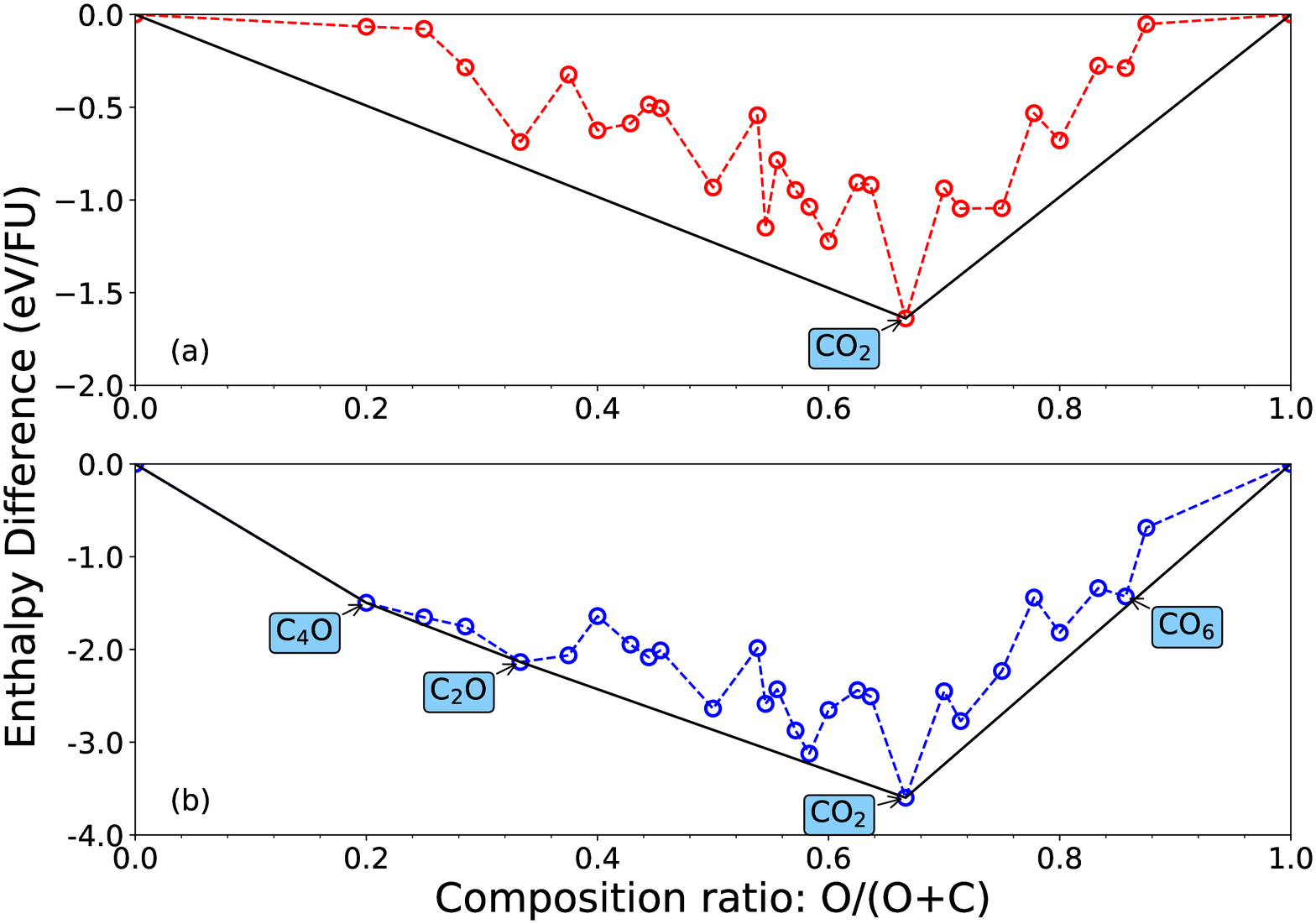}
  \caption{Enthalpy difference per formula unit as a function of composition
    is plotted. (a) Representative enthalpy calculations for varying
    compositions at 7 Mbar. (b) Enthalpy calculations at 50 Mbar suggest C$_4$O and
    C$_2$O structures would be favorable in systems that contain more
    carbon than CO$_2$.}
\label{Fig:COResults}
\end{figure} 

In Fig.~\ref{Fig:COResults}, we plot the enthalpy difference per atom
between various C$_n$O$_m$ compounds that of the endmember phases,
$\Delta H = H_{\rm C_nO_m} - ( nH_{\rm C}+mH_{\rm O} )$. At every
  pressure under consideration, the most stable CO$_2$ structures,
  that we obtained, reproduced previous results~\cite{Lu2013}. The
convex hull at 7 Mbar in Fig.~\ref{Fig:COResults}a shows that no
stable structures are expected to exist besides CO$_2$ and the two
endmembers. However, convex hull diagram at 50 Mbar in
Fig.~\ref{Fig:COResults}b revealed the existence of two new stable
carbon-rich structures with C$_4$O and C$_2$O
compositions. Interestingly, no stable CO structures were found over
the entire pressure range. All CO structures, that we generated, were
found to have a higher enthalpy than a combination of carbon and
CO$_2$. Also none of our oxygen-rich compounds were found to be
stable. Only one structure with a C:O = 1:6 composition came close to
matching the combined enthalpies of pure oxygen and CO$_2$ but was not
found to be stable in the pressure range up to 50 Mbar.

\begin{figure}[h!]
  \includegraphics[width=0.44\textwidth]{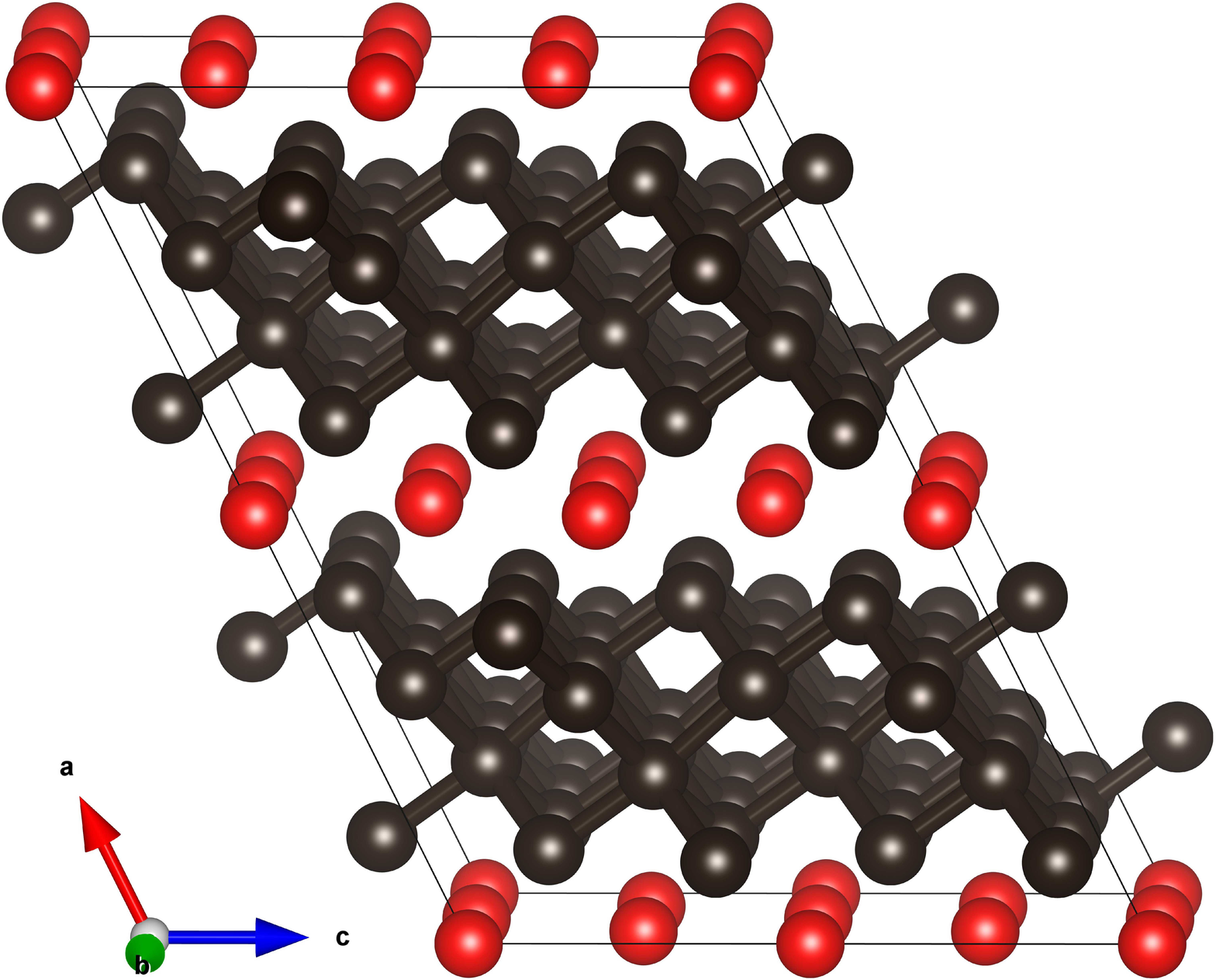}
  \caption{Moniclinic C$_4$O crystal structure with {\it C2/m} symmetry at 45 Mbar. The
    unit cell with 10 atoms as been doubled along every lattice vector
  to better illustrate the C and O layers in the structure. The C
    and O atoms are shown in dark and light color, respectively. }
\label{Fig:imageC4O}
\end{figure}

The C$_4$O structure is monoclinic and has {\em C2/m} symmetry. It was
found from relaxing structures with 2 formula units (10 atoms). The
image in Fig. ~\ref{Fig:imageC4O} reveals a layered structure where
thin oxygen planes alternate with thick carbon layers. The oxygen
atoms form a 2D hexagonal lattice in planes spanned by the crystal
lattice vectors $b$ and $c$. The carbon atoms are arranged on four,
tightly stacked hexagonal layers in between.

\begin{figure}[h!]
  \includegraphics[width=0.45\textwidth]{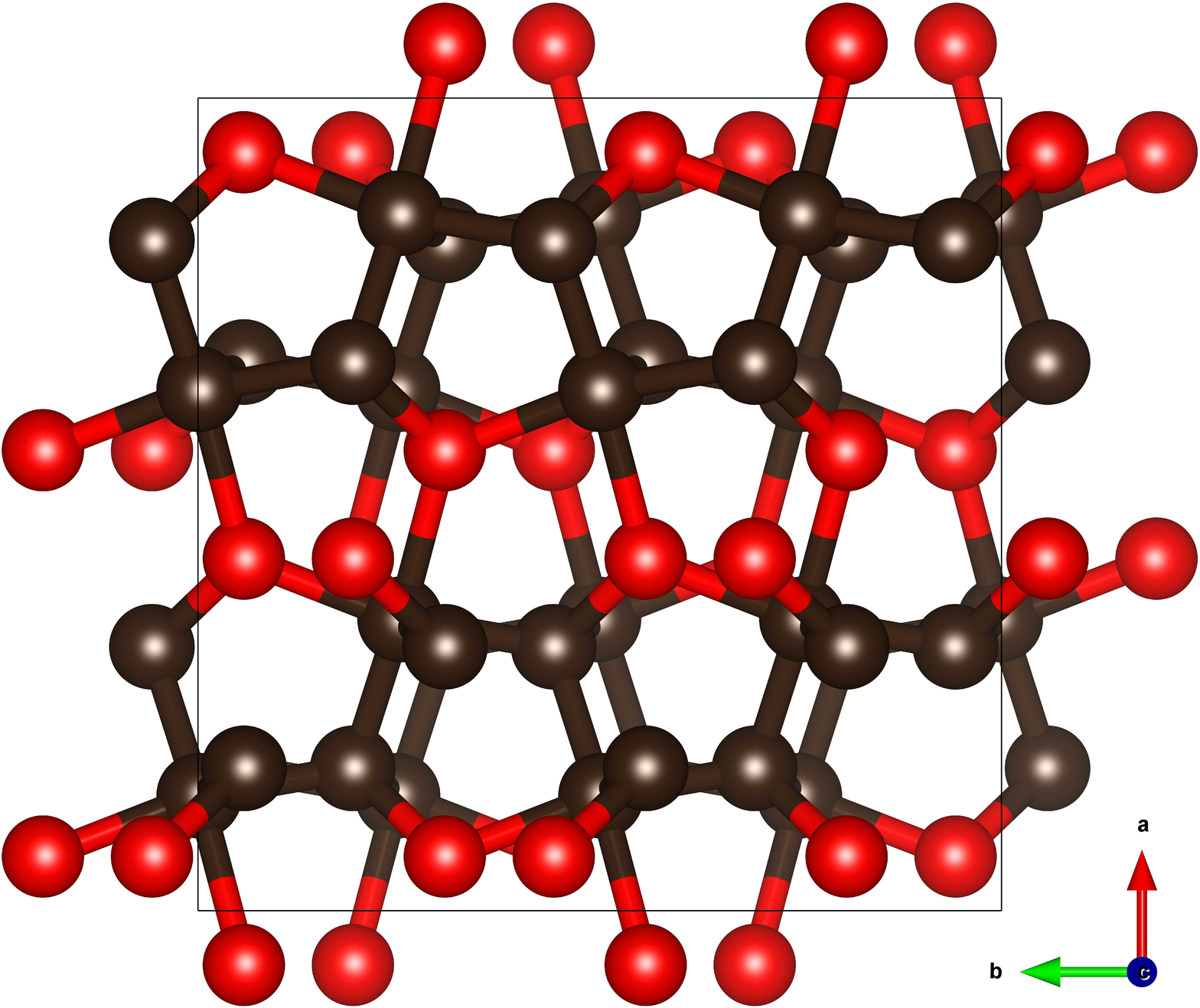}
  \caption{Orthorhombic C$_2$O crystal structure with {\it Pbca} symmetry at 25 Mbar. The
    unit cell with 24 atoms as been doubled in $b$ direction. The C
    and O atoms are shown in dark and light color, respectively.}
\label{Fig:imageC2O}
\end{figure}

The C$_2$O structure can also be viewed as a layered structure but the
bonding is more complex and three dimensional. The structure is
orthorhombic and has {\em Pbca} symmetry. In Fig.~\ref{Fig:imageC2O},
the unit cell with 8 formula units (24 atoms) has been double in $b$
direction to illustrate the layers and 3D bonding. The shortest bonds
occur between the C and O atoms in the layers but C-O bond distances
vary considerably between 1.11 and 1.30 \AA~at 25 Mbar. The C-O layers
are connected by C-C bonds that are all between 1.17 and 1.18
\AA~long. Again, we verified the  C$_2$O and C$_4$O structures structure
  were dynamically stable by performing phonons calculations with the
  Phonopy code~\cite{Togo2015} using 2x2x2 supercells.

\begin{figure}[h]
   \includegraphics[width=.45\textwidth]{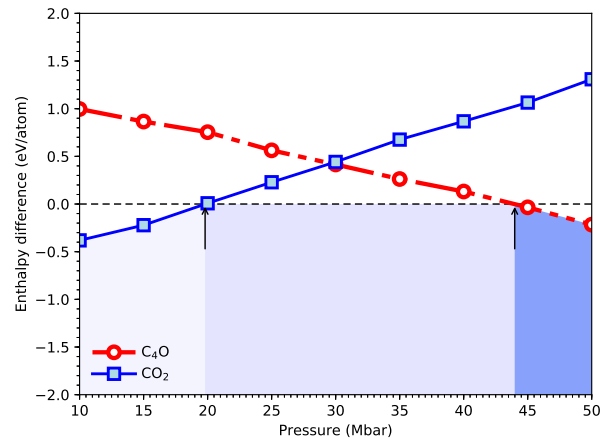}
   \caption{The difference in enthalpy,
     $H_{\rm C_nO_m} - \big[(n-2m) \times H_{\rm C} + m \times H_{\rm C_2O}\big]$, per atom as
     function of pressure. The first arrow marks the pressure of 19.8 Mbar where
     the C$_2$O structure with {\em Pbca} symmetry is predicted to
     form. The second arrow marks the pressure 44.0 Mbar where the C$_4$O
     structure with {\em C2/m} symmetry is predicted to form.}
   \label{Fig:enthalpyCO}
\end{figure}

We performed enthalpy calculations of these new C$_4$O and C$_2$O
structures in order to determine the pressure at which they are
favored over a decomposition into pure carbon and CO$_2$.  In Fig.~\ref{Fig:enthalpyCO}, we choose to plot the resulting enthalpy
difference with respect to a mixture of pure carbon and C$_2$O
because this allows us to illustrate the C$_4$O and the C$_2$O
formulation pressures in a single diagram. We predict the C$_2$O
structure to form at 19.8 Mbar while the C$_4$O structure becomes
stable at 44.0 Mbar. The parameters of both structure given in
tables~\ref{tablePbca} and \ref{tableC2/m}. The formation pressures of
both structures are considerably larger than those that are typically
reached with diamond anvil cell experiments. This is not unexpected
because the diamond anvils would otherwise have reacted with the
samples in any experiment that contained sufficient amounts of free
oxygen. However, such pressures are accessible with dynamic
compression techniques that use ramp waves to compress the sample at
lower temperature than with standard shock wave
experiments~\cite{Eggert2014}.

\begin{table}[h!]
\begin{tabular}{c c c c c}
\hline
\hline
Atom& Wyckoff & x & y    & z \\
\hline
C & c & 0.848   & 0.497 & 0.850 \\
C & c & 0.325   & 0.115 & 0.531\\
O & c & 0.567   & 0.386 & 0.757 \\
\hline
\hline
\end{tabular}
\caption{Parameters of the orthorhombic C$_2$O structure with {\it Pbca} symmetry at 25 Mbar. 
The lattice parameters are $a$=5.845, $b$=2.890, and $c$=2.899 \AA.}
\label{tablePbca}
\end{table}

\begin{table}[h]
\begin{tabular}{c c c c c}
\hline
\hline
Atom& Wyckoff & x & y    & z \\
\hline
C & i & 0.347   & 0.000 & 0.220 \\
C & i & 0.893   & 0.000 & 0.605\\
O & b & 0.000   & 0.500 & 0.000 \\
\hline
\hline
\end{tabular}
\caption{Parameters of the monoclinic C$_4$O structure with {\it C2/m} symmetry at 45 Mbar. The
  lattice parameters are $a$=2.960, $b$=1.434, $c$=3.916 \AA~and $\beta$ = 106.73$^\circ$.}
\label{tableC2/m}
\end{table}

\section{Conclusion}
Our SYDSS algorithm provides a systematic and consistent way to
generate symmetric candidate structures for relaxation with DFT forces
with the goal of predicting novel crystal structures at high
pressure. While no structure is excluded in principle, symmetric
structures are generated with high probability. This significantly
improves the efficiency of our structure search algorithm for large
unit cells with 20 atoms or more, if one adopts the common view that
ground state crystal structures are symmetric.

We applied our SYDSS technique to search for novel stoichiomtric
H$_2$O-NaCl compounds at high pressure because we assumed large unit
cells would be needed to accommodate atoms from both materials in an
optimal way. Indeed, our best structure has a comparatively
large primitive unit cell of 20 atoms.  However, as with any random
search method, there is no guarantee there does not exist yet another
H$_2$O-NaCl structure with lower enthalpy unless our prediction is
confirmed with experiments. The predicted formation pressure of 3.4
Mbar is well within the reach of diamond anvil cell
experiments~\cite{Lobanov2015}. If indeed a yet more stable
H$_2$O-NaCl compounds exists, x-ray diffraction measurements should
reveal such a structure.

When we applied our SYDSS method to search for novel carbon-oxygen
compounds at megabar pressures, we identified two novel carbon-rich
but no oxygen-rich structures. At 19.8 Mbar, we predict an
orthorhombic C$_2$O structure to form from dense carbon and CO$_2$. At
44.0 Mbar, a novel monoclinic C$_4$O structure is expect to become
thermodynamically stable. Both transition pressures are beyond the
reach of static high pressure experiments but can in principle be
generated with dynamic compression techniques.

Also, we cannot completely rule out the existence of unknown
low-enthalpy structures of the H$_2$O, NaCl, carbon, and oxygen
endmembers even though one has looked for such structures carefully
with DFT methods carefully already. If a novel H$_2$O, NaCl, carbon,
or oxygen structure existed, the formation pressures of the predicted
novel compounds would be shifted to higher values than we have predict
here. However, in a diamond cell or ramp compression experiment one
would see such novel endmember structures. In either case, new
compounds or endmember structures are exprected to produced when
H$_2$O-NaCl and C-O mixtures are exposed to pressures of 3.4 and 19.8
Mbar, respectively.

\acknowledgments{The authors acknowledge support from the U.S. National Science Foundation (grant 1412646), the U.S. Department of Energy (grant DE-SC0010517), and University of California's lab fee program as well as encouraging discussions with R.
  Caracas and other participants of the program ``Dynamics and
  Evolution of Earth-like Planets'' at the Kavli Institute for
  Theoretical Physics. In part, this work used the National Energy
  Research Scientific Computing Center and the Extreme Science and
  Engineering Discovery Environment. K. Driver provided comments on this
  manuscript.  }


\end{document}